\documentstyle[12pt]{article} 
\if@twoside \oddsidemargin 21pt \evensidemargin 59pt \marginparwidth 85pt
\else \oddsidemargin 0pt \evensidemargin 0pt
\fi
\newcounter{formel1}

\begin{document}

\title{Interrelations between Stochastic Equations for Systems
with Pair Interactions}

\author{Dirk Helbing \\II. Institute for Theoretical Physics\\
University of Stuttgart, Germany}
\maketitle

\begin{abstract}
Several types of stochastic equations are important in thermodynamics, 
chemistry, evolutionary biology, population dynamics and quantitative 
social science.
For systems with pair interactions four different 
types of equations are derived,
starting from a master equation for the state space:
First, general mean value and (co)variance equations. 
Second, {\sc Boltzmann}-like
equations. Third, a master equation for the configuration space
allowing transition rates which depend on the occupation numbers of the
states. Fourth, a {\sc Fokker-Planck} equation and a 
``{\sc Boltz\-mann-Fokker-Planck} equation''. The interrelations of these
equations and the conditions for their validity
are worked out clearly. 
A procedure for a selfconsistent solution of the nonlinear 
equations is proposed. Generalizations to interactions between an arbitrary
number of systems are discussed.
\end{abstract}
%

\section{Introduction}

Stochastic equations have shown to be a useful tool in all fields of science,
where fluctuations are involved. Many applications can e.g. be found
in chemical kinetics \cite{Opp}, laser theory \cite{Hak} and
biological systems \cite{Arn},
but they also become increasingly important in quantitative social
science \cite{Col,Bar,Wei,Hel,Hel1,Hel2,Hel3,Hel4}. 
Without stochastic models certain phenomena
like {\em phase transitions} could not be correctly understood
\cite{Ma,Hor,Pr}.
\par
Most often,
problems involving fluctuations are handled by master equations,
{\sc Fokker-Planck} equations or {\sc Boltzmann} equations. There are
essentially three ways of arriving at these equations:
\begin{itemize}
\item A system is influenced by {\em external} fluctuations: In this
case, a stochastic differential equation (or {\sc Langevin} equation)
would be set up, which can be transformed to a {\sc Fokker-Planck} equation
\cite{Gardiner,Stra}.
\item A system consists of a huge number of subsystems:
If the entire system behaves conservative, it would be
described classically by a {\sc Liouville} equation or quantum mechanically
by a {\sc von Neumann} equation. Both equations can be contracted to a 
(generalized) master equation 
\cite{Pri,Hov1,Hov2,Res1,Res2,Res3,Mont,Zwanz,Zwan} 
(which becomes an ordinary master
equation in the {\sc Markovian} limit \cite{Ken}) or to a
{\sc Boltzmann} equation 
\cite{Lib,Bogol,Cohen1,Uhl1,Kirk,Grad,Rieckers,Der,Weidl}.
The master equation and the {\sc Boltzmann} equation 
contain a few {\em relevant (macroscopic)} variables only,
and handle the huge remaining number of irrelevant variables as {\em internal}
fluctuations.
\item The state of a system changes to one of several others with a certain
{\em probability} which depends on the actual state ({\sc Markov}
property): This kind of situation would be modelled by a {\sc
Chapman-Kolmogorov} equation, which can be transformed to a master
equation \cite{Haken}.
\end{itemize}
In the following sections the interrelations of master, {\sc Fokker-Planck}
and {\sc Boltzmann} equations shall be discussed
(similar considerations can be made for {\em generalized}
master equations \cite{Risken}, {\em generalized} {\sc Fokker-Planck} equations
\cite{Risken} and {\em generalized} {\sc Boltzmann} 
equations \cite{Der}): Starting from the master
equation for the state space (sect. \ref{MA}), 
the {\sc Boltzmann} equation results in the
case of a factorization of the pair distribution function (sect.
\ref{Ableitung}), whereas the 
{\sc Fokker-Planck} equation follows by a second order
{\sc Taylor} expansion (sect. \ref{FOKK}). In addition, a 
``{\sc Boltzmann-Fokker-Planck} equation'' is derived (sect. \ref{BOL}),
which can be either obtained from the {\sc Fokker-Planck} equation
by a factorization of the pair distribution function or from the
{\sc Boltzmann} equation by a second order {\sc Taylor} expansion.
The nonlinear dependence of this equation on the probability distribution
calls for a selfconsistent method of solution (sect. \ref{SELF}).
\par
Having formulated a master equation for the state space, the same can be done
for the configuration space (occupation numbers' space) 
(sect. \ref{Config}). The mean value equation of the configurational
master equation is equivalent to the {\sc Boltzmann} equation, if simultaneous
interactions of more than two systems are neglected (sect. \ref{Config}).
The {\sc Boltzmann} equation is applicable only during a certain time interval
in which the (co)variances are small (sect. \ref{Config}). 
However, the mean value equation can be corrected by methods, which are also
suitable for calculating the effects of higher order interactions 
(sect. \ref{CORR}).
                                                                         
\section{The master equation} \label{MA}

Ensembles of statistically behaving systems can often be described
by a {\em master equation} \cite{Haken,Kampen,Kampen1}
\begin{equation}
 \frac{d}{dt}P(\vec{X},t) = \int d^l X' \,
\bigg[ w(\vec{X},\vec{X}';t)P(\vec{X}',t) - w(\vec{X}',\vec{X};t)P(\vec{X},t) 
\bigg]
\label{master}
\end{equation}
with
\begin{equation}
 0 \le P(\vec{X},t) \quad \mbox{and}\quad
\int d^l X \, P(\vec{X},t) = 1 \, .
\label{NORM}
\end{equation}
$P(\vec{X},t)$ is the probability of the ensemble to be at time
$t$ in a state $\vec{X} = (x_1,\dots,x_l)$, if a transition from state
$\vec{X}$ to $\vec{X}'$ occurs with probability $w(\vec{X}',\vec{X};t)$
per time unit. The entire state
\begin{eqnarray*}
 \vec{X} & = & (\vec{x}_1;\dots;\vec{x}_\alpha;\dots;\vec{x}_N) \nonumber \\
 & = & (x_{11},\dots,x_{1n};\dots;x_{\alpha 1},\dots,x_{\alpha n};
\dots;x_{N1},\dots,x_{Nn})
\end{eqnarray*}
is combined of the states $\vec{x}_\alpha = (x_{\alpha 1},\dots,
x_{\alpha i},\dots,x_{\alpha n})$
of the ensemble's $N$ individual systems ($l=n\cdot N$).
\par
A typical example for the application of (\ref{master}) is a chemical 
reaction \cite{Opp,Kampen1,Chem1,Chem2,Chem3,Chem4,Chem5,Chem7}
\begin{equation}
 B + C \rightleftharpoons BC \, ,
\label{reaction}
\end{equation}
where $\vec{x}_\alpha$ could, for example, indicate, if atom
$\alpha$ is bound or free. We would have a discrete state space 
with $n=1$ then,
and the integrals in (\ref{master}), (\ref{NORM}) could be replaced by sums:
\begin{displaymath}
\frac{d}{dt}P(\vec{X},t) = \sum_{\vec{X}'}
\bigg[ w(\vec{X},\vec{X}';t)P(\vec{X}',t) - w(\vec{X}',\vec{X};t)P(\vec{X},t) 
\bigg] \, ,
\end{displaymath}
\begin{displaymath}
 \sum_{\vec{X}} P(\vec{X},t) = 1 \, .
\end{displaymath}
Methods for solving master equations are discussed in
\cite{m1,m2,m3,Mont,master1,master2,master3,master4,master5,Kampen1,Weid1,Weid2}.

\subsection{Mean value equations (Macroscopic equations)} \label{meanvalues}

Usually, one is mainly interested in the time evolution of the mean value
\begin{displaymath}
\langle f \rangle := \int d^l X \, f(\vec{X},t)P(\vec{X},t) 
\end{displaymath}
of $f(\vec{X},t) = x_{\alpha i}$. This is given by \cite{Kampen1}
\begin{eqnarray}
 \frac{d \langle x_{\alpha i} \rangle}{d t} 
&=& \int d^l X\, x_{\alpha i} \frac{dP(\vec{X},t)}{dt} \nonumber \\
&=& \int d^l X \int d^l X'\, \bigg[x_{\alpha i} w(\vec{X},\vec{X}';t)
P(\vec{X}',t) - x_{\alpha i} w(\vec{X}',\vec{X};t)P(\vec{X},t)
\bigg] \nonumber \\
&=&  \int d^l X \int d^l X'\, (x'_{\alpha i} - x_{\alpha i}) 
w(\vec{X}', \vec{X};t)P(\vec{X},t) \nonumber \\
&=&  \int d^l X\, m_{\alpha i}(\vec{X},t) P(\vec{X},t) \nonumber \\
&=& \langle m_{\alpha i}(\vec{X},t) \rangle
\label{mean1}
\end{eqnarray}
with the {\em first jump moments} \cite{Kampen1}
\begin{equation}
 m_{\alpha i}(\vec{X},t) := \int d^l X' \, \Delta x'_{\alpha i}
w(\vec{X}',\vec{X};t)
\qquad (\Delta x'_{\alpha i} := x'_{\alpha i} - x_{\alpha i}) \, .
\label{moment1}
\end{equation}
In the derivation of (\ref{mean1}) we have exchanged the order of
$\vec{X}$ and $\vec{X}'$. 
\par
If the probability distribution 
$P(\vec{X},t)$ has only small (co)variances, we find in first order 
{\sc Taylor} approximation the relation \cite{Kampen1}
\begin{equation}
\frac{d \langle x_{\alpha i} \rangle}{d t} 
\approx \bigg\langle m_{\alpha i}(\langle \vec{X}\rangle,t)
+ \sum_{\beta j} (x_{\beta j} - \langle x_{\beta j} \rangle)
\frac{\partial m_{\alpha i}(\langle \vec{X}\rangle,t)}{\partial 
\langle x_{\beta j} \rangle} \bigg\rangle
= m_{\alpha i}(\langle \vec{X} \rangle,t)  \, .
\label{approx1}
\end{equation}
In many cases, the initial state $\vec{X}_0$ at a time $t_0$ is known
by measurement or preparation,
i.e. the initial distribution is
\begin{displaymath}
 P(\vec{X},t_0) = \delta(\vec{X}-\vec{X}_0)
\end{displaymath}
with the {\sc Dirac} delta function $\delta(.)$. As a consequence,
the (co)variances 
\begin{displaymath}
\sigma_{\alpha i \beta j} 
:= \bigg\langle (x_{\alpha i}-\langle x_{\alpha i} \rangle)
(x_{\beta j} - \langle x_{\beta j} \rangle) \bigg\rangle
= \langle x_{\alpha i}x_{\beta j} \rangle 
- \langle x_{\alpha i} \rangle\langle x_{\beta j} \rangle 
\end{displaymath}
vanish at time $t_0$ and remain small during a certain time interval.
Proceeding as above, the equations
\begin{equation}
 \frac{d \sigma_{\alpha i \beta j}}{d t}
=\langle m_{\alpha i \beta j}(\vec{X},t) \rangle
+ \bigg\langle ( x_{\alpha i} - \langle x_{\alpha i}\rangle) 
m_{\beta j}(\vec{X},t)\bigg\rangle 
+ \bigg\langle (x_{\beta j}- \langle x_{\beta j} \rangle )
 m_{\alpha i}(\vec{X},t) \bigg\rangle 
\label{mean2}
\end{equation}
for the temporal development of the 
(co)variances are found \cite{Kampen1}, where
\begin{equation}
 m_{\alpha i\beta j}(\vec{X},t) 
:= \int d^l X' \,  \Delta x'_{\alpha i} \Delta x'_{\beta j}
w(\vec{X}', \vec{X};t) 
\label{moment2}
\end{equation}
are the {\em second jump moments}.
The (co)variance equations (\ref{mean2}) have, in first order 
{\sc Taylor} approximation
\begin{eqnarray*}
& &  \bigg\langle ( x_{\alpha i} - \langle x_{\alpha i}\rangle) f(\vec{X},t)
\bigg\rangle \nonumber \\
&\approx& \left\langle  (x_{\alpha i} 
- \langle x_{\alpha i}\rangle)f(\langle \vec{X} \rangle ,t) +
 ( x_{\alpha i} - \langle x_{\alpha i}\rangle) \sum_{\gamma,k}
( x_{\gamma k} - \langle x_{\gamma k}\rangle) \frac{\partial 
f(\langle \vec{X} \rangle,t)}{\partial \langle x_{\gamma k} \rangle}
\right\rangle \nonumber \\
&=& \sum_{\gamma,k} \sigma_{\alpha i\gamma k}\frac{\partial 
f(\langle \vec{X} \rangle,t)}{\partial \langle x_{\gamma k} \rangle} \, ,
\end{eqnarray*}
the form \cite{Kampen1}
\begin{equation}
\frac{d\sigma_{\alpha i\beta j}}{d t} 
\approx m_{\alpha i \beta j}(\langle \vec{X} \rangle,t)
+\sum_{\gamma, k}\left( 
\sigma_{\alpha i \gamma k} \frac{\partial m_{\beta j}(\langle
\vec{X}\rangle,t)}{\partial \langle x_{\gamma k} \rangle} + 
\sigma_{\beta j\gamma k} \frac{\partial m_{\alpha i}(\langle \vec{X}
\rangle,t)}{\partial \langle x_{\gamma k} \rangle} \right) \, .
\label{approx2}
\end{equation}
(Since equations (\ref{mean1}) and (\ref{mean2}) are no closed equations,
they can only be approximately solved.)

\subsection{{\sc Boltzmann}-like equations} \label{Ableitung}

The probability distribution $P(\vec{X},t)$
rather than the mean value equations should be calculated,
if the (co)variances are not negligible. Since 
the number $N$ of systems is usually tremendous, (\ref{master})
can be solved neither exactly nor by computer simulations. In order to find
a suitable simplification of (\ref{master}), let us introduce the vectors
\begin{displaymath}
 \vec{X}^\alpha := (\vec{0};\dots;\vec{0};\vec{x}_\alpha;\vec{0};
\dots;\vec{0}) \, ,
\end{displaymath}
\begin{displaymath}
 \vec{X}^{\alpha_1\dots\alpha_k} := \sum_{i=1}^k \vec{X}^{\alpha_i} \, ,
\end{displaymath}
and let us decompose the transition rates
\begin{eqnarray}
 w(\vec{Y};\vec{X};t) & = &\sum_{k=1}^N 
\sum_{\alpha_1<\dots<\alpha_k}
w_{\alpha_1\dots\alpha_k}\left(\vec{Y}^{\alpha_1\dots\alpha_k};
\vec{X}^{\alpha_1\dots\alpha_k};t\right) \nonumber \\
&=& \sum_{k=1}^N \frac{1}{k!}
\sum_{\alpha_1,\dots,\alpha_k}
w_{\alpha_1\dots\alpha_k}\left(\vec{Y}^{\alpha_1\dots\alpha_k};
\vec{X}^{\alpha_1\dots\alpha_k};t\right)
\label{decompose}
\end{eqnarray}
into terms
\begin{displaymath}
 w_{\alpha_1\dots\alpha_k}\left(\vec{Y}^{\alpha_1\dots\alpha_k};
\vec{X}^{\alpha_1\dots\alpha_k};t\right)
\end{displaymath}
which describe the interactions between exactly $k$ systems. 
Consequently, we have to set
\begin{displaymath}
w_{\alpha_1\dots\alpha_k}
\equiv 0, \quad \mbox{if two indices $\alpha_i$, $\alpha_j$ agree.}
\end{displaymath}
\par
For most of the chemical reactions and many other cases the terms
with $k\le 2$ are the only important ones, because simultaneous interactions
between three or more systems are rare. Therefore, the terms with
$k >2$ shall be neglected in the following, but a generalization to 
an arbitrary number of simultaneously interacting systems is possible 
(see sect. \ref{more}) \cite{Bogol,Cohen1,Uhl,Cohen2,Hanley}.
The master equation (\ref{master}) has then the form
\begin{eqnarray}
 \frac{dP(\vec{X},t)}{dt} 
&=& \sum_\beta \int \! d^l Y \, \bigg[ w_\beta(\vec{X}^\beta;
\vec{Y}^\beta;t)
P(\vec{Y},t) 
- w_\beta(\vec{Y}^\beta;\vec{X}^\beta;t)P(\vec{X},t) \bigg]
\nonumber \\
&+& \frac{1}{2} \sum_{\beta, \gamma}\!\!
\int \! d^l Y \, \bigg[ w_{\beta\gamma}(\vec{X}^{\beta\gamma};
\vec{Y}^{\beta\gamma};t)
P(\vec{Y},t) - w_{\beta\gamma}
(\vec{Y}^{\beta\gamma};\vec{X}^{\beta\gamma},t)
P(\vec{X},t) \bigg] .
\label{analog}
\end{eqnarray}
Let us use
\begin{equation}
 \int d^n x_\beta \, P_{_{\alpha_1\dots\beta\dots\alpha_k}}
(\vec{x}_{\alpha_1};\dots;\vec{x}_\beta;\dots;\vec{x}_{\alpha_k};t) 
= P_{_{\alpha_1\dots\alpha_k}}
(\vec{x}_{\alpha_1};\dots;\vec{x}_{\alpha_k};t)
\label{integration}
\end{equation}
with
\begin{displaymath}
 P_{_{1\dots N}} 
(\vec{x}_1;\dots;\vec{x}_N;t)
\equiv P(\vec{x}_1;\dots;\vec{x}_N;t) \, ,
\end{displaymath}
and assume the factorization
\begin{equation}
P_{\alpha\beta}(\vec{x}_\alpha;\vec{x}_\beta;t) 
\approx P_\alpha(\vec{x}_\alpha,t)P_\beta(\vec{x}_\beta,t) 
\label{factor}
\end{equation}
of the pair distribution $P_{\alpha\beta}(\vec{x}_\alpha;\vec{x}_\beta;t)$.
Integration of (\ref{analog}) over all
variables $\vec{x}_\beta$ with $\beta \ne \alpha$ leads to 
\cite{Kubo,Rieckers}
\begin{eqnarray}
\frac{d}{dt}P_\alpha(\vec{x}_\alpha,t)
&=&\int \!\! d^n y_\alpha \, 
\bigg[ w_\alpha(\vec{x}_\alpha,
\vec{y}_\alpha;t)
P_\alpha(\vec{y}_\alpha,t) - w_\alpha(\vec{y}_\alpha,
\vec{x}_\alpha;t)P_\alpha(\vec{x}_\alpha,t) \bigg]
\nonumber \\
&+& \int \!\! d^n y_\alpha \left[ \left(\sum_{\beta}
\int \!\! d^n y_\beta \int \!\! d^n x_\beta \, 
w_{\alpha\beta}(\vec{x}_\alpha,\vec{x}_\beta ;
\vec{y}_\alpha,\vec{y}_\beta;t)
P_\beta(\vec{y}_\beta,t)\right)P_\alpha(\vec{y}_\alpha,t) 
 \right. \nonumber \\
&&\mbox{ } \left.
-\left(\sum_{\beta} 
\int \!\! d^n y_\beta \int \!\! d^n x_\beta \,  
w_{\alpha \beta}(\vec{y}_{\alpha},\vec{y}_\beta;
\vec{x}_\alpha,\vec{x}_\beta;t)
P_\beta(\vec{x}_\beta,t) \right)P_\alpha(\vec{x}_\alpha,t) \right] 
\label{Boltz}
\end{eqnarray}
with
\begin{equation}
 \int d^n  x_{\alpha}\, P_{\alpha}(\vec{x}_{\alpha},t) = 1 \, ,
\label{Norm}
\end{equation}
where the conventions
\begin{eqnarray*}
w_\alpha(\vec{y}_\alpha,\vec{x}_\alpha;t) &\equiv &
w_\alpha(\vec{Y}^\alpha;\vec{X}^\alpha;t) \, ,\nonumber \\ 
w_{\alpha\beta}(\vec{y}_\alpha,\vec{y}_\beta;\vec{x}_\alpha,
\vec{x}_\beta;t) &\equiv & w_{\alpha\beta}(\vec{Y}^{\alpha\beta};
\vec{X}^{\alpha\beta};t) 
\end{eqnarray*}
have been used.
The second and third terms of (\ref{Boltz}) are similar to those of the 
{\sc Boltzmann} equation \cite{Kubo,Boltz,Kampen1}.
$w_\alpha(\vec{y}_\alpha,\vec{x}_\alpha;t)$ describes transitions of system
$\alpha$ from state $\vec{x}_\alpha$\linebreak 
to state $\vec{y}_\alpha$ occuring spontaneously
or being induced by external influences.\linebreak 
$w_{\alpha\beta}(\vec{y}_\alpha,
\vec{y}_\beta;\vec{x}_\alpha,\vec{x}_\beta;t)$ represents interactions between
two systems $\alpha$ and $\beta$ changing their states from
$\vec{x}_\alpha$ resp. $\vec{x}_\beta$ to $\vec{y}_\alpha$ resp.
$\vec{y}_\beta$. More clearly arranged, (\ref{Boltz}) can be written
in the form
\begin{equation}
 \frac{d}{dt}P_\alpha(\vec{x},t)
=\int \!\! d^n x' \, 
\bigg[ w^\alpha(\vec{x},
\vec{x}';t)
P_\alpha(\vec{x}',t) - w^\alpha(\vec{x}',
\vec{x};t)P_\alpha(\vec{x},t) \bigg]
\label{Boltz2}
\end{equation}
with the {\em effective} transition rates
\begin{equation}
 w^\alpha(\vec{x}',\vec{x};t) :=
w_\alpha(\vec{x}',\vec{x};t) + \sum_{\beta}
\int d^n y' \int d^n y \,  
w_{\alpha \beta}(\vec{x}',\vec{y}';
\vec{x},\vec{y};t)
P_\beta(\vec{y},t) \, .
\label{abbr2}
\end{equation}
The factorization (\ref{factor}) is valid, if
each two systems $\alpha$ and $\beta$ are statistically independent. This is
approximately the case in the limit of weak interactions ($w_{\alpha \beta}
\approx 0$) or, if correlations between systems $\alpha$ and $\beta$
due to pair interactions are soon destroyed, e.g. 
by external influences
(see also (\ref{Small})). Equation
(\ref{factor}) is called the assumption of {\em molecular chaos}
\cite{Keizer}. Without this assumption, equation (\ref{Boltz})
would contain the unknown pair distribution $P_{\alpha\beta}
(\vec{x}_\alpha,\vec{x}_\beta,t)$. However, there are methods of
calculating corrections, if the factorization (\ref{factor}) is not
valid \cite{Bogol,Dorf} (see sect. \ref{more}). 

\subsection{Types of interaction} \label{Types}

Normally, the huge number $N$ of systems $\alpha$ can be divided into $A$
{\em types} (resp. sets) $a$ which can be characterized by a special kind of 
interaction. In example (\ref{reaction}), one type would consist of
molecules $B$ and a second type of molecules $C$. 
Let $N_a$ denote the number of systems $\alpha\in a$ (i.e. of type $a$).
This implies
\begin{displaymath}
 \sum_{a=1}^A N_a = N \, .
\end{displaymath}
If we assume systems
$\alpha \in a$ of the same type $a$ to be indistinguishable, we have 
\begin{displaymath}
 P_\alpha \equiv P_a \quad \mbox{if} \quad 
\alpha \in a \, , 
\end{displaymath}
\begin{displaymath}
 w_\alpha \equiv \widetilde{w}_{a} \quad \mbox{if} \quad \alpha \in a \, , 
\end{displaymath}
\begin{displaymath}
 w_{\alpha\beta} \equiv \left\{ \begin{array}{ll}
\widetilde{w}_{ab} &\mbox{if } \alpha \ne \beta, \mbox{ }
 \alpha \in a, \mbox{ } \beta \in b \\
0 &\mbox{if } \alpha = \beta \, ,
\end{array} \right.
\end{displaymath}
where $\widetilde{w}_{..}$ are {\em individual} transition rates.
We may now introduce 
the abbreviations
\begin{eqnarray*}
 w_a &:=& \widetilde{w}_a  \, , \\
 w_{ab} &:=& \left\{
\begin{array}{ll}
 N_b\cdot \widetilde{w}_{ab} & \mbox{if } b\ne a \\
 (N_a -1)\cdot \widetilde{w}_{ab}   
 & \mbox{if } b=a 
\,.
\end{array} \right. \nonumber
\end{eqnarray*}
Then, from (\ref{Boltz2}), (\ref{abbr2}), (\ref{Norm}) we get
\begin{equation}
\frac{d}{dt}P_a(\vec{x},t) =\int \!\! d^n x' \, 
\bigg[ w^a(\vec{x},
\vec{x}';t)
P_a(\vec{x}',t) - w^a(\vec{x}',
\vec{x};t)P_a(\vec{x},t) \bigg]
\label{Boltz1}
\end{equation}
with the effective transition rates
\begin{equation}
 w^a(\vec{x}',\vec{x};t) :=
w_a(\vec{x}',\vec{x};t) + \sum_{b}
\int d^n y' \int d^n y \,  
w_{ab}(\vec{x}',\vec{y}';
\vec{x},\vec{y};t)
P_b(\vec{y},t) 
\label{abbrev}
\end{equation}
and
\begin{displaymath}
 \int d^n x \, P_a (\vec{x},t) = 1 \, .
\end{displaymath}
In equations (\ref{Boltz1}), (\ref{abbrev}) 
the number of variables has been drastically reduced (if $A\ll N$ resp.
$N_a \gg 1$). As
a consequence, (\ref{Boltz1}), (\ref{abbrev}) 
are tractable with analytical or computer methods.
Especially, for $A=1$, $w_1 \equiv 0$,
\begin{displaymath}
 \int d^6 x' \int d^6 y' \, w_{11}(\vec{x}',\vec{y}';\vec{x},\vec{y})
= \int d^6 x' \int d^6 y' \, w_{11}(\vec{x},\vec{y};\vec{x}',\vec{y}')
\end{displaymath}
and
\begin{displaymath}
 \vec{x} := (\vec{r},\vec{v}) \, ,
\end{displaymath}
we obtain the ordinary {\sc Boltzmann} equation for gases
\cite{Landau,Boltz}, where 
$\vec{r}$ denotes the place and $\vec{v}$ the velocity of a particle. 
A recursive method for solving this special case has been
given by {\sc Chapman} \cite{Chapman,Chapman1} and {\sc Enskog} \cite{Enskog}.

\subsection{The configuration and its temporal development} \label{Config}

In many cases, we are not interested in the states $\vec{x}_\alpha$
of the single systems $\alpha$, but in the {\em number} $n_{\vec{x}}^a$
of systems of type $a$ being in state $\vec{x}$
(e.g. the number of molecules of type $B$ 
being 
bound). Let us assume that there is only
a finite number $S$ of possible states $\vec{x}\in \{\vec{x}_1,\dots,
\vec{x}_S \}$ (although analogous considerations can be made 
for a continuum of states $\vec{x}$).
Then, the vector
\begin{displaymath}
 \vec{n} := (n_{\vec{x}_1}^1,\dots,
n_{\vec{x}_S}^1;\dots;n_{\vec{x}_1}^A,\dots,n_{\vec{x}_S}^A) 
 = (\vec{n}^1;\dots;\vec{n}^A)
\end{displaymath}
of occupation numbers $n_{\vec{x}}^a$
shall be called the {\em configuration} of the ensemble of systems.
The configuration space consists of all 
vectors $\vec{n}$ obeying the relations
\begin{displaymath}
\sum_{\vec{x}} n_{\vec{x}}^a \equiv \sum_{s=1}^S n_{\vec{x}_s}^a = N_a \, .
\end{displaymath}
If we neglect pair interactions for a moment
(i.e. if we only take terms of order $k=1$, $w_{ab} \equiv 0$), 
we can assume the individual systems to be
statistically independent from each other. The probability distribution of the
configurations $\vec{n}$ will then be
\begin{equation}
 P(\vec{n},t) = \prod_{a=1}^A P^a(\vec{n}^a,t)
\label{Pe}
\end{equation}
with the multinomial distribution \cite{Poisson}
\begin{equation}
 P^a(\vec{n}^a,t) \equiv
 P^a(n_{\vec{x}_1}^a,\dots,n_{\vec{x}_S}^a;t)
= \frac{N_a !}{n_{\vec{x}_1}^a!\cdot\dots\cdot n_{\vec{x}_S}^a!}
 P_a(\vec{x}_1,t)^{n_{\vec{x}_1}^a}\cdot \dots \cdot 
P_a(\vec{x}_S,t)^{n_{\vec{x}_S}^a} \, .
\label{Poisson}
\end{equation}
By differentiation of (\ref{Pe}), (\ref{Poisson})
and use of (\ref{Boltz1}), we obtain for the temporal change of $P(\vec{n},t)$
the master equation 
\begin{equation}
 \frac{d}{dt} P(\vec{n},t)
= \sum_{\vec{n}'} \bigg[ w(\vec{n},\vec{n}';t)P(\vec{n}',t)
- w(\vec{n}',\vec{n};t)P(\vec{n},t) \bigg] 
\label{master2}
\end{equation}
with
\begin{displaymath}
0 \le P(\vec{n},t) \, , \qquad  \sum_{\vec{n}} P(\vec{n},t) = 1 
\end{displaymath}
and
\begin{equation}
 w(\vec{n}',\vec{n};t) = \left\{
\begin{array}{ll}
n_{\vec{x}}^a \widetilde{w}_a(\vec{y},\vec{x};t) & \mbox{if } 
\vec{n}' = \vec{n}^a_{\vec{y}\vec{x}} \\
0 & \mbox{else\,,}
\end{array} \right. 
\label{we}
\end{equation}
\begin{displaymath}
 \vec{n}^a_{\vec{y}\vec{x}} := (n_{\vec{x}_1}^1,\dots,n_{\vec{x}_S}^1;\dots;
n_{\vec{x}_1}^a,\dots,
(n_{\vec{y}}^a +1),\dots,(n_{\vec{x}}^a - 1),\dots,
n_{\vec{x}_S}^a;\dots;n_{\vec{x}_1}^A,\dots,n_{\vec{x}_S}^A) \, .
\end{displaymath}
According to this, the {\em configurational} transition rate
$n_{\vec{x}}^a \widetilde{w}_a(\vec{y},\vec{x};t)
=n_{\vec{x}}^a w_a(\vec{y},\vec{x};t)$
is proportional to the number $n_{\vec{x}}^a$ of systems which can change
the state $\vec{x}$ and proportional to the {\em individual} transition rate
$\widetilde{w}_a(\vec{y},\vec{x};t)$.
\par
The equations for the mean values 
\begin{displaymath}
 \langle f \rangle := \sum_{\vec{n}} f(\vec{n},t)P(\vec{n},t)
\end{displaymath}
of $f(\vec{n},t) = n_{\vec{x}}^a$ 
can be obtained from (\ref{mean1}) and the first jump moments
\begin{eqnarray}
 m_{\vec{x}}^a (\vec{n},t)  &=& \sum_{\vec{n}'}
(\Delta \vec{n}')_{\vec{x}}^a w(\vec{n}',\vec{n};t) \nonumber \\
&=& \sum_{\vec{y},\vec{y}',b}
(\Delta \vec{n}_{\vec{y}\vec{y}'}^b)_{\vec{x}}^a
w(\vec{n}_{\vec{y}\vec{y}'}^b,\vec{n};t) 
\nonumber \\
&=& \sum_{\vec{y}} \left[n_{\vec{y}}^a w_a(\vec{x},\vec{y};t)
- n_{\vec{x}}^a w_a(\vec{y},\vec{x};t) \right] \, .
\label{first}
\end{eqnarray}
(Here, $(\vec{n})_{\vec{x}}^a$ denotes the component $n_{\vec{x}}^a$
of $\vec{n}$.)
After substitution of $\vec{y}$ by $\vec{x}'$, the 
mean value equations of (\ref{master2}), (\ref{we}) have the 
explicit form
\begin{equation}
 \frac{d}{dt} \langle n_{\vec{x}}^a \rangle = 
\langle m_{\vec{x}}^a(\vec{n},t) \rangle = \sum_{\vec{x}'}
\bigg[w_a(\vec{x},\vec{x}';t)\langle n_{\vec{x}'}^a \rangle
- w_a (\vec{x}',\vec{x};t)\langle n_{\vec{x}}^a \rangle \bigg] \, .
\label{mean}
\end{equation}
Therefore, the equations governing the temporal change
of the expected fractions $\langle n_{\vec{x}}^a \rangle/N_a$
agree with (\ref{Boltz1}), 
and we have:
\begin{equation}
 \frac{\langle n_{\vec{x}}^a\rangle}{N_a}= P_a(\vec{x},t) \, .
\label{frequencies}
\end{equation}

\subsubsection*{Pair interactions}

In order to take pair interactions into account, we apply the method of
section \ref{Ableitung}. The master equation (\ref{master2}) then
reads
\begin{eqnarray}
 \frac{d}{dt} P(\vec{n},t) &=& \sum_{a,\vec{x},a',\vec{x}'}
\bigg[ w(\vec{n};\vec{n}_{\vec{x}'}^{a'}{}_{\vec{x}}^a;t)
P(\vec{n}_{\vec{x}'}^{a'}{}_{\vec{x}}^a,t) 
- w(\vec{n}_{\vec{x}'}^{a'}{}_{\vec{x}}^a;\vec{n};t)P(\vec{n},t) \bigg]
\nonumber \\
&+& \frac{1}{2}\sum_{a,\vec{x},a',\vec{x}'}\sum_{b,\vec{y},b',\vec{y}'}
\bigg[ w(\vec{n};
\vec{n}_{\vec{x}'}^{a'}{}_{\vec{y}'}^{b'}{}_{\vec{x}}^a{}_{\vec{y}}^b;t)
P(\vec{n}_{\vec{x}'}^{a'}{}_{\vec{y}'}^{b'}{}_{\vec{x}}^a{}_{\vec{y}}^b,t)
\nonumber \\
& & \qquad \qquad
- w(\vec{n}_{\vec{x}'}^{a'}{}_{\vec{y}'}^{b'}{}_{\vec{x}}^a{}_{\vec{y}}^b;
\vec{n};t)P(\vec{n},t) \bigg]
\label{Mastereq}
\end{eqnarray}
with
\begin{displaymath}
 \vec{n}_{\vec{x}'}^{a'}{}_{\vec{x}}^a
:= (\dots,(n_{\vec{x}'}^{a'}+1),\dots,(n_{\vec{x}}^a-1),\dots) \, ,
\end{displaymath}
\begin{displaymath}
 \vec{n}_{\vec{x}'}^{a'}{}_{\vec{y}'}^{b'}{}_{\vec{x}}^a{}_{\vec{y}}^b
:= (\dots,(n_{\vec{x}'}^{a'}+1),\dots,
(n_{\vec{x}}^a -1),\dots,(n_{\vec{y}'}^{b'}+1),\dots,
(n_{\vec{y}}^b -1),\dots) \, .
\end{displaymath}
Usually, the transition rates will depend on the changing occupation
numbers only, which results in
\begin{eqnarray*}
 \frac{d}{dt} P(\vec{n},t) &=& \sum_{a,\vec{x},a',\vec{x}'}
\bigg[ w_{\vec{x}}^a{}_{\vec{x}'}^{a'}
(n_{\vec{x}}^a-1,n_{\vec{x}'}^{a'}+1;t)
P(\vec{n}_{\vec{x}'}^{a'}{}_{\vec{x}}^a,t) 
- w_{\vec{x}'}^{a'}{}_{\vec{x}}^a
(n_{\vec{x}'}^{a'},n_{\vec{x}}^{a};t)P(\vec{n},t) \bigg]
\nonumber \\
&+& \frac{1}{2}\sum_{a,\vec{x},a',\vec{x}'}\sum_{b,\vec{y},b',\vec{y}'}
\bigg[ w_{\vec{x}}^{a}{}_{\vec{y}}^{b}{}_{\vec{x}'}^{a'}{}_{\vec{y}'}^{b'}
(n_{\vec{x}}^a-1,n_{\vec{y}}^b-1;n_{\vec{x}'}^{a'}+1,n_{\vec{y}'}^{b'}+1;t)
P(\vec{n}_{\vec{x}'}^{a'}{}_{\vec{y}'}^{b'}{}_{\vec{x}}^a{}_{\vec{y}}^b,t)
\nonumber \\
& &\qquad \qquad \quad 
- w_{\vec{x}'}^{a'}{}_{\vec{y}'}^{b'}{}_{\vec{x}}^a{}_{\vec{y}}^b
(n_{\vec{x}'}^{a'},n_{\vec{y}'}^{b'};n_{\vec{x}}^{a},n_{\vec{y}}^{b};t)
P(\vec{n},t) \bigg] 
\end{eqnarray*}
analogous to (\ref{analog}).
Let us in the following consider the special case
\begin{eqnarray*}
 w_{\vec{x}'}^{a'}{}_{\vec{x}}^a
(n_{\vec{x}'}^{a'},n_{\vec{x}}^{a};t)
&:=& n_{\vec{x}}^a \widetilde{w}_a(\vec{x}',\vec{x};t)\delta_{aa'} \nonumber \\
&=&  n_{\vec{x}}^a w_a(\vec{x}',\vec{x};t)\delta_{aa'} \, , \\
w_{\vec{x}'}^{a'}{}_{\vec{y}'}^{b'}{}_{\vec{x}}^a{}_{\vec{y}}^b
(n_{\vec{x}'}^{a'},n_{\vec{y}'}^{b'};n_{\vec{x}}^{a},n_{\vec{y}}^{b};t)
&:=& n_{\vec{x}}^a n_{\vec{y}}^b \widetilde{w}_{ab}(\vec{x}',\vec{y}';
\vec{x},\vec{y};t) \delta_{aa'} \delta_{bb'} \nonumber \\
&\stackrel{N_a \gg 1}{\approx}&  
n_{\vec{x}}^a \frac{n_{\vec{y}}^b}{N_b} w_{ab}(\vec{x}',\vec{y}';
\vec{x},\vec{y};t) \delta_{aa'} \delta_{bb'} \, ,
\end{eqnarray*}
that means
\begin{equation}
 w(\vec{n}',\vec{n};t) = \left\{
\begin{array}{ll}
n_{\vec{x}}^a \widetilde{w}_a(\vec{x}',\vec{x};t) & \mbox{if } 
\vec{n}' = \vec{n}_{\vec{x}'}^a{}_{\vec{x}}^a \\
n_{\vec{x}}^a n_{\vec{y}}^b \widetilde{w}_{ab}(\vec{x}',\vec{y}';
\vec{x},\vec{y};t) & \mbox{if } \vec{n}' 
=\vec{n}_{\vec{x}'}^{a}{}_{\vec{y}'}^{b}{}_{\vec{x}}^a{}_{\vec{y}}^b \\
0 & \mbox{else}
\end{array} \right.
\label{Raten}
\end{equation}
(where $\delta_{aa'}$ denotes the {\sc Kronecker} function).
The configurational transition rates are proportional to the
individual transition rates $\widetilde{w}_a$ resp. $\widetilde{w}_{ab}$
and proportional to the numbers $n_{\vec{x}}^a$ 
of systems which can change the state $\vec{x}$ resp. to the numbers
$n_{\vec{x}}^a n_{\vec{y}}^b$ of possible pair interactions between
states $\vec{x}$ and $\vec{y}$. Therefore, if $\vec{n}' 
=\vec{n}_{\vec{x}'}^{a}{}_{\vec{y}'}^{a}{}_{\vec{x}}^a{}_{\vec{x}}^a$,
$n_{\vec{x}}^a n_{\vec{x}}^a \widetilde{w}_{aa}(\vec{x}',\vec{y}';
\vec{x},\vec{x};t)$ has to be replaced by $n_{\vec{x}}^a (n_{\vec{x}}^a -1)
\widetilde{w}_{aa}(\vec{x}',\vec{y}';\vec{x},\vec{x};t)$ in order to
exclude self-interactions again. However, (\ref{Raten}) is approximately
valid, if $n_{\vec{x}}^a \gg 1$
for all configurations $\vec{n}$, where $P(\vec{n},t)$ cannot be neglected.
For simplicity, the validity of this assumption will be also presupposed 
in the following sections.
%

\subsubsection*{Mean value equations (Macroscopic equations)}

Noticing 
\begin{displaymath}
 w_{\beta\alpha}(\vec{y}',\vec{x}';\vec{y},\vec{x};t)
= w_{\alpha\beta}(\vec{x}',\vec{y}';\vec{x},\vec{y};t)
\end{displaymath}
(since the transition rates are independent from the arbitrary 
numeration of the systems), the first jump moments can be calculated
analogously to equation (\ref{first}). They are found to be
\begin{equation}
 m_{\vec{x}}^a (\vec{n},t)  
= \sum_{\vec{x}'} \left[n_{\vec{x}'}^a \overline{w}^a(\vec{x},\vec{x}';t)
- n_{\vec{x}}^a \overline{w}^a(\vec{x}',\vec{x};t) \right] 
\label{mom1}
\end{equation}
with
\begin{equation}
 \overline{w}^a(\vec{x}',\vec{x};t) :=
w_a(\vec{x}',\vec{x};t) + \sum_{b}
\sum_{\vec{y}'} \sum_{\vec{y}} w_{ab}(\vec{x}',\vec{y}';
\vec{x},\vec{y};t)
\frac{n_{\vec{y}}^b}{N_b} 
\label{noname}
\end{equation}
(if $N_a \gg 1$).
Therefore, the approximate mean value equations (see (\ref{approx1})) are
\begin{equation}
 \frac{d}{dt} \langle n_{\vec{x}}^a \rangle 
\approx m_{\vec{x}}^a (\langle \vec{n} \rangle,t) = \sum_{\vec{x}'}
\bigg[\widehat{w}^a(\vec{x},\vec{x}';t)
\langle n_{\vec{x}'}^a \rangle
-\widehat{w}^a (\vec{x}',\vec{x};t)
\langle n_{\vec{x}}^a \rangle \bigg] 
\label{Mean}
\end{equation}
with the {\em mean transition rates}
\begin{equation}
 \widehat{w}^a(\vec{x}',\vec{x};t) :=
w_a(\vec{x}',\vec{x};t) + \sum_{b}
\sum_{\vec{y}'} \sum_{\vec{y}} w_{ab}(\vec{x}',\vec{y}';
\vec{x},\vec{y};t)
\frac{\langle n_{\vec{y}}^b \rangle}{N_b} \, .
\label{Abbrev}
\end{equation}
Again, equations (\ref{Mean}), (\ref{Abbrev}) for the expected
fractions $\langle n_{\vec{x}}^a \rangle/N_a$
agree with equations (\ref{Boltz1}), (\ref{abbrev}) for the
probabilities $P_a(\vec{x},t)$ if making the identification 
(\ref{frequencies}). However, the approximate equations (\ref{Mean}),
(\ref{Abbrev}) are only valid under the condition
\begin{displaymath}
 \langle n_{\vec{x}}^a n_{\vec{y}}^b \rangle
 \approx  \langle n_{\vec{x}}^a \rangle \langle n_{\vec{y}}^b \rangle \, ,
\end{displaymath}
which obviously corresponds to the factorization assumption (\ref{factor}). 
This condition is fulfilled if the absolute values of the (co)variances
\begin{displaymath}
 \sigma_{\vec{x}}^a{}_{\vec{x}'}^b =
 \langle n_{\vec{x}}^a n_{\vec{x}'}^b \rangle 
 - \langle n_{\vec{x}}^a \rangle\langle n_{\vec{x}'}^b \rangle 
\end{displaymath}
are small, i.e., if
\begin{equation}
\left|  \sigma_{\vec{x}}^a{}_{\vec{x}'}^b \right| \ll
 \langle n_{\vec{x}}^a \rangle\langle n_{\vec{x}'}^b \rangle \, .
 \label{Small}
\end{equation}
(This is often the case in the thermodynamic limes $N\longrightarrow \infty$).
According to (\ref{approx2}), approximate equations for the
temporal development of the (co)variances are
\begin{equation}
 \frac{d \sigma_{\vec{x}}^a{}_{\vec{x}'}^b}{d t}
 \approx m_{\vec{x}}^a{}_{\vec{x}'}^b(\langle \vec{n} \rangle,t)
+ \sum_{c,\vec{y}} \left(
\sigma_{\vec{x}}^a{}_{\vec{y}}^c
\frac{\partial m_{\vec{x}'}^b (\langle \vec{n}\rangle,t)}{\partial
\langle n_{\vec{y}}^c \rangle}
+\sigma_{\vec{x}'}^b{}_{\vec{y}}^c
\frac{\partial m_{\vec{x}}^a(\langle \vec{n} \rangle,t)}{\partial
\langle n_{\vec{y}}^c \rangle} \right)
\label{Var}
\end{equation}
with the second jump moments
\begin{eqnarray}
m_{\vec{x}}^a{}_{\vec{x}'}^b (\vec{n},t) &=& \sum_{\vec{n}'}
(\Delta \vec{n}')_{\vec{x}}^a (\Delta \vec{n}')_{\vec{x}'}^b 
w(\vec{n}',\vec{n};t) \nonumber \\
&=& \sum_{\vec{y},\vec{y}',c}
(\Delta \vec{n}_{\vec{y}}^c{}_{\vec{y}'}^c)_{\vec{x}}^a
(\Delta \vec{n}_{\vec{y}}^c{}_{\vec{y}'}^c)_{\vec{x}'}^b
w(\vec{n}_{\vec{y}}^c{}_{\vec{y}'}^c, \vec{n};t) \nonumber \\
&+& \frac{1}{2} \sum_{\vec{y},\vec{y}',c} \sum_{\vec{z},\vec{z}',d}
(\Delta 
\vec{n}_{\vec{y}'}^c{}_{\vec{z}'}^d{}_{\vec{y}}^c{}_{\vec{z}}^d)_{\vec{x}}^a
(\Delta 
\vec{n}_{\vec{y}'}^c{}_{\vec{z}'}^d{}_{\vec{y}}^c{}_{\vec{z}}^d)_{\vec{x}'}^b 
w(\vec{n}_{\vec{y}'}^c{}_{\vec{z}'}^d{}_{\vec{y}}^c{}_{\vec{z}}^d;\vec{n};t)
\nonumber \\
&=& \delta_{ab} \bigg( \delta_{\vec{x}\vec{x}'}
\sum_{\vec{y}}\bigg[ n_{\vec{y}}^a \overline{w}^a(\vec{x},\vec{y};t)
+ n_{\vec{x}}^a \overline{w}^a(\vec{y},\vec{x};t) \bigg] \nonumber \\
& & \qquad  - \bigg[n_{\vec{x}'}^a \overline{w}^a(\vec{x},\vec{x}';t)
+  n_{\vec{x}}^a \overline{w}^a(\vec{x}', \vec{x};t) \bigg]\bigg) \nonumber \\
&+& \sum_{\vec{y}'}\sum_{\vec{y}} \bigg[ 
n_{\vec{y}}^a \frac{n_{\vec{y}'}^b}{N_b} w_{ab}
(\vec{x},\vec{x}';\vec{y},\vec{y}';t) 
+ n_{\vec{x}}^a \frac{n_{\vec{x}'}^b}{N_b} w_{ab}
(\vec{y},\vec{y}';\vec{x},\vec{x}';t) \bigg] \nonumber \\
&-& \sum_{\vec{y}'}\sum_{\vec{y}} \bigg[
n_{\vec{x}}^a \frac{n_{\vec{y}'}^b}{N_b}  w_{ab}
(\vec{y},\vec{x}';\vec{x},\vec{y}';t) 
+ n_{\vec{y}}^a\frac{n_{\vec{x}'}^b}{N_b} w_{ab}
(\vec{x},\vec{y}';\vec{y},\vec{x}';t) \bigg] \, .
\label{mom2}
\end{eqnarray}
The approximate mean value equations (\ref{Mean}), (\ref{mom1}), (\ref{noname})
and the approximate (co)variance equations (\ref{Var}), (\ref{mom2})
can be generalized to the case of transition rates
\begin{equation}
 w(\vec{n}',\vec{n};t) = \left\{
\begin{array}{ll}
n_{\vec{x}}^a \widetilde{w}_a(\vec{x}',n_{\vec{x}'}^a,
\vec{x},n_{\vec{x}}^a;t) & \mbox{if } 
\vec{n}' = \vec{n}_{\vec{x}'}^a{}_{\vec{x}}^a \\
n_{\vec{x}}^a n_{\vec{y}}^b \widetilde{w}_{ab}(\vec{x}',n_{\vec{x}'}^a,\vec{y}',
n_{\vec{y}'}^b;\vec{x},n_{\vec{x}}^a,\vec{y},n_{\vec{y}}^b;t) 
& \mbox{if } \vec{n}' 
=\vec{n}_{\vec{x}'}^{a}{}_{\vec{y}'}^{b}{}_{\vec{x}}^a{}_{\vec{y}}^b \\
0 & \mbox{else.}
\end{array} \right. \label{DEPEND}
\end{equation}
This generalization is necessary, if
the individual transition rates $\widetilde{w}_a$, $\widetilde{w}_{ab}$
are themselves functions of the occupation numbers. Such cases appear
e.g. in quantitative social science, where the decisions or the behavior
of individuals may depend on the {\em socio}configuration $\vec{n}$ 
\cite{Wei,Hel1,Weid1,Weid2,Weid3,Weid4}. Equations (\ref{Mastereq}) resp.
(\ref{Mean}), (\ref{mom1}), (\ref{noname}) and (\ref{Var}), (\ref{mom2})
can also be used for {\em mean field} approaches in
physics, e.g the {\sc Ising} model \cite{Ash,Ising}. 

\subsection{Corrections and higher order interactions} \label{CORR}
\label{more}
In section \ref{Config} the exact mean value equations
\begin{equation}
 \frac{d\langle n_{\vec{x}}^a\rangle}{dt}
= \langle m_{\vec{x}}^a (\vec{n},t) \rangle
\label{MEAN1}
\end{equation}
are, according to equations (\ref{mom1}), (\ref{noname}) 
for $m_{\vec{x}}^a$, dependent
on $\langle n_{\vec{x}}^a n_{\vec{y}}^b\rangle$. However, the exact
equation
\begin{displaymath}
 \frac{d\langle n_{\vec{x}}^a n_{\vec{x}'}^b\rangle}{dt}
= \langle m_{\vec{x}}^a{}_{\vec{x}'}^b(\vec{n},t) \rangle
+ \langle n_{\vec{x}}^a m_{\vec{x}'}^b(\vec{n},t) \rangle
+ \langle n_{\vec{x}'}^b m_{\vec{x}}^a(\vec{n},t) \rangle
\end{displaymath}
governing the temporal development of 
$\langle n_{\vec{x}}^a n_{\vec{x}'}^b\rangle$ depends on
$\langle n_{\vec{x}}^a n_{\vec{x}'}^b n_{\vec{x}'{}'}^c \rangle$,
because of equation (\ref{mom2}). So, we are confronted with the problem
of having no closed set of equations. 
\par
Let us consider the general case of up to $k$ interactions. In the 
master equation (\ref{master2}) we have then to take the transition
rates 
\begin{equation}
 w(\vec{n}',\vec{n};t) = \left\{
\begin{array}{ll}
n_{\vec{x}_1}^{a_1} \widetilde{w}_{a_1}^{a'_1}(\vec{x}'_1,\vec{x}_1;t) 
& \mbox{if } \vec{n}' = \vec{n}_{\vec{x}'_1}^{a'_1}{}_{\vec{x}_1}^{a_1} \\
n_{\vec{x}_1}^{a_1} n_{\vec{x}_2}^{a_2} 
\widetilde{w}_{a_1}^{a'_1}{}_{a_2}^{a'_2}
(\vec{x}'_1,\vec{x}'_2;\vec{x}_1,\vec{x}_2;t) & \mbox{if } \vec{n}' 
=\vec{n}_{\vec{x}'_1}^{a'_1}{}_{\vec{x}'_2}^{a'_2}{}
_{\vec{x}_1}^{a_1}{}_{\vec{x}_2}^{a_2} \\
\quad \vdots & \quad \vdots \\
n_{\vec{x}_1}^{a_1}\cdot\dots\cdot n_{\vec{x}_k}^{a_k}
\widetilde{w}_{a_1}^{a'_1}{}_{\dots}^{\dots}{}_{a_k}^{a'_k}
(\vec{x}'_1,\dots,\vec{x}'_k;\vec{x}_1,\dots,\vec{x}_k;t)
& \mbox{if } \vec{n}' = \vec{n}_{\vec{x}'_1}^{a'_1}{}_{\dots}^{\dots}{}
_{\vec{x}'_k}^{a'_k}{}_{\vec{x}_1}^{a_1}{}_{\dots}^{\dots}{}
_{\vec{x}_k}^{a_k} \\
0 & \mbox{else}
\end{array} \right.
\label{Trans}
\end{equation}
with
\begin{displaymath}
 \vec{n}_{\vec{x}'_1}^{a'_1}{}_{\dots}^{\dots}{}
_{\vec{x}'_k}^{a'_k}{}_{\vec{x}_1}^{a_1}{}_{\dots}^{\dots}{}
_{\vec{x}_k}^{a_k}
:= (\dots,(n_{\vec{x}'_1}^{a'_1}+1),\dots,(n_{\vec{x}_1}^{a_1}-1),
\dots,(n_{\vec{x}'_k}^{a'_k}+1),\dots,(n_{\vec{x}_k}^{a_k} -1),\dots) \, .
\end{displaymath}
The temporal development of the $l$th moments
$\langle n_{\vec{x}_1}^{a_1}\cdot\dots\cdot n_{\vec{x}_l}^{a_l}\rangle$
is governed by the equations
\begin{eqnarray}
 \frac{d}{d t}
\langle n_{\vec{x}_1}^{a_1}\cdot\dots\cdot n_{\vec{x}_l}^{a_l}\rangle
&=& \sum_{\vec{n}}\sum_{\vec{n}'}
\bigg[ (\vec{n}')_{\vec{x}_1}^{a_1}\cdot\dots\cdot(\vec{n}')_{\vec{x}_l}^{a_l}
- n_{\vec{x}_1}^{a_1}\cdot\dots\cdot n_{\vec{x}_l}^{a_l}\bigg]
w(\vec{n}',\vec{n};t)P(\vec{n},t) \nonumber \\
&=& \sum_{\vec{n}}\sum_{\vec{n}'} \bigg[
\bigg( (\Delta \vec{n}')_{\vec{x}_1}^{a_1} + n_{\vec{x}_1}^{a_1} \bigg)
\cdot\dots\cdot\bigg( (\Delta \vec{n}')_{\vec{x}_l}^{a_l}
+ n_{\vec{x}_l}^{a_l}\bigg) \nonumber \\
& & \qquad - n_{\vec{x}_1}^{a_1}\cdot\dots\cdot n_{\vec{x}_l}^{a_l}\bigg]
w(\vec{n}',\vec{n};t)P(\vec{n},t) \nonumber \\
&=& \sum_{m=1}^l \sum_{{\cal M}_m}
\left\langle m_{\vec{x}_{i_1}}^{a_{i_1}}{}_{\dots}^{\dots}{}_{\vec{x}_{i_m}}
^{a_{i_m}} \cdot n_{\vec{x}_{i_{m+1}}}^{a_{i_{m+1}}}\cdot\dots\cdot 
n_{\vec{x}_{i_l}}^{a_{i_l}}\right\rangle \nonumber \\
&=& \sum_{m=1}^l \frac{1}{m!(l-m)!} \sum_{(i_1,\dots,i_l)\in \sigma(l)}
\left\langle m_{\vec{x}_{i_1}}^{a_{i_1}}{}_{\dots}^{\dots}{}_{\vec{x}_{i_m}}
^{a_{i_m}} \cdot n_{\vec{x}_{i_{m+1}}}^{a_{i_{m+1}}}\cdot\dots\cdot 
n_{\vec{x}_{i_l}}^{a_{i_l}}\right\rangle \, .
\label{MEANS}
\end{eqnarray}
Here, 
\begin{displaymath}
 m_{\vec{x}_{i_1}}^{a_{i_1}}{}_{\dots}^{\dots}{}_{\vec{x}_{i_m}}^{a_{i_m}}
(\vec{n},t)
:= \sum_{\vec{n}'}(\Delta \vec{n}')_{\vec{x}_{i_1}}^{a_{i_1}}\cdot\dots\cdot
(\Delta \vec{n}')_{\vec{x}_{i_m}}^{a_{i_m}}w(\vec{n}',\vec{n};t)
\end{displaymath}
are the $m$th jump moments, and $\sigma(l)$ is the set of all $l!$ permutations
of the indices $(1,\dots,l)$. $\sum_{{\cal M}_m}$ means the summation over
all subsets $\{i_1,\dots,i_m\}$ of $\{1,\dots,l\}\equiv\{i_1,\dots,i_l\}$
consisting of $m$ elements. Notice, that the solutions for the moments should,
in the case of time independent transition rates $w(\vec{n}',\vec{n})$,
converge to certain values in the limit $t\longrightarrow \infty$, since 
$P(\vec{n},t)$ approaches an unique equilibrium distribution 
\cite{Gardiner}.
\par
Because of (\ref{Trans}),
the $m$th jump moments are polynomials of order $k$ of the
occupation numbers $n_{\vec{x}_{i_n}}^{a_{i_n}}$.
Therefore, the equations (\ref{MEANS}) for the $l$th moments depend on
the 1st up to the \mbox{$(l-1+k)$th} 
moments. As a consequence, the exact moment
equations (\ref{MEANS}) represent a closed system of equations only
for $k=1$ (exclusion of 
multiple interactions). For interactions of up
to $k>1$ systems we have to take a suitable approximation of (\ref{MEANS})
or, in other words, of
\begin{eqnarray}
& & n_{\vec{x}_1}^{a_1}\cdot\dots\cdot n_{\vec{x}_n}^{a_n} \nonumber \\
&=& \bigg[\bigg(n_{\vec{x}_1}^{a_1} - \langle n_{\vec{x}_1}^{a_1}\rangle
\bigg) + \langle n_{\vec{x}_1}^{a_1} \rangle \bigg]\cdot\dots\cdot
\bigg[\bigg(n_{\vec{x}_n}^{a_n} - \langle n_{\vec{x}_n}^{a_n}\rangle
\bigg) + \langle n_{\vec{x}_n}^{a_n} \rangle \bigg] \label{Exact} \\
&=& \sum_{m=0}^n \sum_{{\cal M}_m}
\bigg(n_{\vec{x}_{i_1}}^{a_{i_1}} - \langle
n_{\vec{x}_{i_1}}^{a_{i_1}} \rangle\bigg) \cdot \dots \cdot 
\bigg(n_{\vec{x}_{i_m}}
^{a_{i_m}} - \langle n_{\vec{x}_{i_m}}^{a_{i_m}} \rangle\bigg)
\bigg\langle n_{\vec{x}_{i_{m+1}}}^{a_{i_{m+1}}} \bigg\rangle 
\cdot \dots \cdot
\bigg\langle n_{\vec{x}_{i_l}}^{a_{i_l}} \bigg\rangle \, .
\label{EXACT}
\end{eqnarray}
(\ref{EXACT}) is the {\em exact} {\sc Taylor} expansion of
(\ref{Exact}). 
If we are interested in the 1st up to the $j$th 
moments ($1 \le l\le j$, $j\ge 1$), we have essentially
two possibilities of getting approximate moment equations:
\begin{enumerate}
\item[(a)] Replacing all jump moments 
$m_{\vec{x}_{i_1}}^{a_{i_1}}{}_{\dots}^{\dots}{}_{\vec{x}_{i_m}}^{a_{i_m}}$
by first order {\sc Taylor} approximations:
\begin{displaymath}
  m_{\vec{x}_{i_1}}^{a_{i_1}}{}_{\dots}^{\dots}{}_{\vec{x}_{i_m}}^{a_{i_m}}
(\vec{n},t) \approx 
 m_{\vec{x}_{i_1}}^{a_{i_1}}{}_{\dots}^{\dots}{}_{\vec{x}_{i_m}}^{a_{i_m}}
(\langle \vec{n} \rangle,t) 
+ \sum_{a,\vec{x}} (n_{\vec{x}}^a - \langle n_{\vec{x}}^a \rangle)
\frac{\partial}{\partial \langle n_{\vec{x}}^a \rangle}
 m_{\vec{x}_{i_1}}^{a_{i_1}}{}_{\dots}^{\dots}{}_{\vec{x}_{i_m}}^{a_{i_m}}
(\langle\vec{n}\rangle,t) \, .
\end{displaymath}
Then, the equations for the $l$th moments depend on 
the 1st up to the $l$th moments only, but each equation is an approximation.
This method has been used in sections \ref{meanvalues} and \ref{Config}.
According to (\ref{EXACT}), it corresponds to the assumption
\begin{displaymath}
\left| \left\langle (n_{\vec{x}_{i_1}}^{a_{i_1}} - \langle
n_{\vec{x}_{i_1}}^{a_{i_1}} \rangle) \cdot \dots \cdot (n_{\vec{x}_{i_m}}
^{a_{i_m}} - \langle n_{\vec{x}_{i_m}}^{a_{i_m}} \rangle)\right\rangle
\right| \ll  \left\langle n_{\vec{x}_{i_1}}^{a_{i_1}} \right\rangle 
\cdot \dots \cdot \left\langle n_{\vec{x}_{i_m}}^{a_{i_m}} 
\right\rangle 
\quad \mbox{for} \quad 1<m\le k \, ,
\label{APP1}
\end{displaymath}
which can be checked by considering the 1st up to the $k$th moments
(i.e. $j\stackrel{!}{\ge} k$).
\item[(b)] 
Replacing $\Big(m_{\vec{x}_{i_1}}^{a_{i_1}}{}_{\dots}^{\dots}{}_{\vec{x}_{i_m}}
^{a_{i_m}} \cdot n_{\vec{x}_{i_{m+1}}}^{a_{i_{m+1}}}\cdot\dots\cdot 
n_{\vec{x}_{i_l}}^{a_{i_l}}\Big)$
by $j$th order {\sc Taylor} approximations. Then, the
equations for the 1st up to the $(j-k+1)$th moments are exact
(see (\ref{EXACT})), but the equation for an $l$th moment
depends on the 1st up to the $(l-1+k)$th moments, some of which may
be determined by approximate equations. Method (b) corresponds to the
assumption
\begin{equation}
\left| \left\langle (n_{\vec{x}_{i_1}}^{a_{i_1}} - \langle
n_{\vec{x}_{i_1}}^{a_{i_1}} \rangle) \cdot \dots \cdot (n_{\vec{x}_{i_m}}
^{a_{i_m}} - \langle n_{\vec{x}_{i_m}}^{a_{i_m}} \rangle)\right\rangle
\right| \ll  \left\langle n_{\vec{x}_{i_1}}^{a_{i_1}} \right\rangle 
\cdot \dots \cdot \left\langle n_{\vec{x}_{i_m}}^{a_{i_m}} 
\right\rangle \mbox{ for } j < m \le l - 1 + k 
\label{LIMES}
\end{equation}
with $1 \le l \le j$.
For large $m$
(\ref{LIMES}) is normally fulfilled:
If the (co)variances are small in the sense of 
\begin{equation}
\left| \sigma_{\vec{x}}^a{}_{\vec{x}'}^b \right| < 
\langle n_{\vec{x}}^a \rangle \langle n_{\vec{x}'}^b \rangle
\label{sense}
\end{equation}
(which is often true in the thermodynamic limes
$N\longrightarrow \infty$), the absolute value of 
\begin{displaymath}
\left(\frac{n_{\vec{x}_{i_1}}^{a_{i_1}}}{\langle
n_{\vec{x}_{i_1}}^{a_{i_1}} \rangle}-1\right) 
\cdot \dots \cdot \left(\frac{n_{\vec{x}_{i_m}}
^{a_{i_m}}}{\langle n_{\vec{x}_{i_m}}^{a_{i_m}} \rangle} - 1\right)
\end{displaymath}
is usually small where $P(\vec{n},t)$ is large \cite{Gardiner}. Note, that
(\ref{sense}) has to be checked only for the variances 
$\sigma_{\vec{x}}^a{}_{\vec{x}}^a$, since the {\sc Cauchy-Schwarz} inequality
implies
\begin{displaymath}
\left| \sigma_{\vec{x}}^a{}_{\vec{x}'}^b \right| \le
\sqrt{\sigma_{\vec{x}}^a{}_{\vec{x}}^a\sigma_{\vec{x}'}^b{}_{\vec{x}'}^b}\, .
\end{displaymath}
\end{enumerate}
As higher order approximation, method
(b) will usually yield better results than
(a) if $j\ge 2$. 
For $k=2$ (i.e. pair interactions) and $j=2$ it would e.g. result in 
corrected mean value equations (\ref{MEAN1}), (\ref{mom1}), (\ref{noname})
containing terms of the form $\langle n_{\vec{x}}^a n_{\vec{x}'}^b\rangle$.
That means, the temporal development of the macroscopic quantities
$\langle n_{\vec{x}}^a \rangle$ depend on the {\em fluctuations}
(resp. the variances $\sigma_{\vec{x}}^a{}_{\vec{x}'}^b
= \langle n_{\vec{x}}^a n_{\vec{x}'}^b \rangle - \langle n_{\vec{x}}^a \rangle
\langle n_{\vec{x}'}^b \rangle$)! The
corrected (co)variance equations resulting from method (b) would be
\begin{displaymath}
 \frac{d \sigma_{\vec{x}}^a{}_{\vec{x}'}^b}{d t}
 \approx m_{\vec{x}}^a{}_{\vec{x}'}^b(\langle \vec{n} \rangle,t)
+ \frac{1}{2}\sum_{c,\vec{y}}\sum_{d,\vec{y}'} 
\sigma_{\vec{y}}^c{}_{\vec{y}'}^d
\frac{\partial^2 m_{\vec{x}}^a{}_{\vec{x}'}^b(\langle \vec{n}\rangle,t)}
{\partial \langle n_{\vec{y}}^c \rangle\partial\langle n_{\vec{y}'}^d\rangle}
+ \sum_{c,\vec{y}} \left(
\sigma_{\vec{x}}^a{}_{\vec{y}}^c
\frac{\partial m_{\vec{x}'}^b (\langle \vec{n}\rangle,t)}{\partial
\langle n_{\vec{y}}^c \rangle}
+\sigma_{\vec{x}'}^b{}_{\vec{y}}^c
\frac{\partial m_{\vec{x}}^a(\langle \vec{n} \rangle,t)}{\partial
\langle n_{\vec{y}}^c \rangle} \right) \, ,
\end{displaymath}
using
\begin{displaymath}
 \langle n_{\vec{x}_1}^{a_1}n_{\vec{x}_2}^{a_2}n_{\vec{x}_3}^{a_3}\rangle
\approx \langle n_{\vec{x}_1}^{a_1}\rangle\langle n_{\vec{x}_2}^{a_2}\rangle
\langle n_{\vec{x}_3}^{a_3}\rangle 
+ \sigma_{\vec{x}_1}^{a_1}{}_{\vec{x}_2}^{a_2}\langle
n_{\vec{x}_3}^{a_3}\rangle
+ \sigma_{\vec{x}_1}^{a_1}{}_{\vec{x}_3}^{a_3}\langle
n_{\vec{x}_2}^{a_2}\rangle
+ \sigma_{\vec{x}_2}^{a_2}{}_{\vec{x}_3}^{a_3}\langle
n_{\vec{x}_1}^{a_1}\rangle 
\end{displaymath}
and (\ref{mom2}).
\par
Methods (a) and (b) are also applicable in the case of configuration dependent
individual transition rates (see (\ref{DEPEND})).

\section{The {\sc Fokker-Planck} equation} \label{FOKK}
 
The master equation (\ref{master}) can 
(by substitution of $\vec{Z}:= - \Delta \vec{X}' \equiv \vec{X}-\vec{X}'$
in the first term and of $\vec{Z} := +\Delta \vec{X}'$ in the second term)
be written in the form
\begin{eqnarray}
\frac{d}{dt}P(\vec{X},t) & = &  \int d^l Z \,  \bigg[
w(\vec{X},\vec{X}-\vec{Z};t) P(\vec{X}-\vec{Z},t) -
w(\vec{X}+\vec{Z}, \vec{X};t)P(\vec{X},t) \bigg] \nonumber \\
& \equiv & \int d^l Z \, \bigg[
w[\vec{X}-\vec{Z};\vec{Z};t] P(\vec{X}-\vec{Z},t) -
w[\vec{X};\vec{Z};t]P(\vec{X},t) \bigg]
\label{Master}
\end{eqnarray}
with
\begin{displaymath}
w[\vec{X};\vec{Z};t] := w(\vec{X}+\vec{Z}, \vec{X};t) 
\qquad (\vec{Z}\equiv\Delta \vec{X}')\, . 
\end{displaymath}
According to {\sc Kramers} \cite{Kramers}
and {\sc Moyal} \cite{Moyal}, equation (\ref{Master})
may be transformed
into the {\sc Fokker-Planck} equation \cite{Gardiner}
\begin{equation}
 \frac{d}{dt}P(\vec{X},t) =
- \sum_{\alpha, i} \frac{\partial}{\partial x_{\alpha i}} 
\bigg[m_{\alpha i}(\vec{X},t)P(\vec{X},t)\bigg]
+ \frac{1}{2} \sum_{\alpha, i}\sum_{\beta, j} 
\frac{\partial}{\partial x_{\alpha i}}\frac{\partial}
{\partial x_{\beta j}} \bigg[m_{\alpha i \beta j}(\vec{X},t)P(\vec{X},t)
\bigg] 
\label{Planck}
\end{equation}
by a second order {\sc Taylor} approximation.
This approximation will be the better the smaller the (co)variances
$\sigma_{\alpha i\beta j}$ are (see the {\sc van Kampen} system size
expansion \cite{Kampen2,Kampen1,Gardiner} 
for a more systematic derivation). However, the 
{\sc Fokker-Planck} equation implies the same mean value and (co)variance
equations 
as the master equation \cite{Weid1}, and is compatible with
\begin{displaymath}
 0 \le P(\vec{X},t) \, , \qquad \int d^l X \, P(\vec{X},t) = 1 
\end{displaymath}
(see the {\sc Pawula} theorem \cite{Risken}).
Interpreting equation (\ref{Planck}), $m_{\alpha i}$ (see (\ref{moment1}))
have the meaning of {\em drift coefficients} and $m_{\alpha i \beta j}$
(see (\ref{moment2})) the meaning of {\em diffusion coefficients}
\cite{Risken,Kampen1}.
\par
The {\sc Fokker-Planck} equation corresponding to the master equation
(\ref{master2}), (\ref{Raten}) is
\begin{equation}
 \frac{d}{dt}P(\vec{n},t)
= -\sum_{a,\vec{x}} \frac{\partial}{\partial n_{\vec{x}}^a}
\bigg[m_{\vec{x}}^a(\vec{n},t)P(\vec{n},t) \bigg]
+\frac{1}{2}\sum_{a,\vec{x}}\sum_{b,\vec{x}'}
\frac{\partial}{\partial n_{\vec{x}}^a}\frac{\partial}{\partial 
n_{\vec{x}'}^b} \bigg[
m_{\vec{x}}^a{}_{\vec{x}'}^b(\vec{n},t)P(\vec{n},t) \bigg] \, .
\label{FOK}
\end{equation}
(Equation (\ref{FOK}) should be transformed to variables 
$u_{\vec{x}}^a := n_{\vec{x}}^a/N_a$. These variables can be approximately
handled as continuous ones; see \cite{Weid1}.)
\par
A reason for using the {\sc Fokker-Planck} equation instead of the master
equation is the variety of methods existing for its solution 
\cite{Risken,Moss,Gardiner,Fok1,Fok2,Fok3,Fok4,Fok5},
since it is a partial differential equation. Especially, the
{\sc Fokker-Planck} equation has some analogies to the {\sc Schr\"odinger}
equation \cite{Risken,Moss}.

\subsection{The ``{\sc Boltzmann-Fokker-Planck} equation''} \label{BOL}

We shall now proceed with equation (\ref{Planck})
as in section \ref{Ableitung}. Let us decompose
the transition rates as in equation (\ref{decompose}) and neglect 
interactions of $k>2$ systems, again. We find
\begin{eqnarray*}
m_{\alpha i}(\vec{X},t) &=& 
m_{\alpha i}^\alpha (\vec{X}^\alpha,t) + \sum_{\beta} 
m_{\alpha i}^{\alpha \beta}(\vec{X}^{\alpha\beta},t) \, ,
\nonumber \\
m_{\alpha i \beta j}(\vec{X},t) 
&=& \delta_{\alpha \beta} \left[
m_{\alpha i \alpha j}^\alpha (\vec{X}^{\alpha},t)
+ \sum_{\gamma}m_{\alpha i \alpha j}^{\alpha\gamma} 
(\vec{X}^{\alpha\gamma},t) \right]
+ m_{\alpha i \beta j}^{\alpha \beta}(\vec{X}^{\alpha\beta},t) 
\end{eqnarray*}
with 
\begin{eqnarray*}
m_{\alpha i}^\gamma (\vec{X}^\gamma,t)
&:=&\int d^l Z \, z_{\alpha i}w_\gamma[\vec{X}^\gamma;\vec{Z}^\gamma;t] \, ,
\nonumber \\
m_{\alpha i}^{\gamma\delta}(\vec{X}^{\gamma\delta},t)
&:=&\int d^l Z \, z_{\alpha i}w_{\gamma\delta}[\vec{X}^{\gamma\delta}
;\vec{Z}^{\gamma\delta};t]
\, , \nonumber \\
m_{\alpha i \beta j}^\gamma (\vec{X}^\gamma,t)
&:=&\int d^l Z \, z_{\alpha i}z_{\beta j}
w_\gamma[\vec{X}^\gamma;\vec{Z}^\gamma;t] \, , \nonumber \\
m_{\alpha i \beta j}^{\gamma\delta}(\vec{X}^{\gamma\delta},t)
&:=&\int d^l Z \, z_{\alpha i}z_{\beta j}
w_{\gamma\delta}[\vec{X}^{\gamma\delta};\vec{Z}^{\gamma\delta};t] \, . 
\end{eqnarray*}
By integration over all variables $\vec{x}_\beta$ with $\beta \ne \alpha$
(see (\ref{integration})) and assumption of the factorization (\ref{factor}),
one obtains
\begin{displaymath}
 \frac{d}{dt}P_\alpha(\vec{x}_\alpha,t) =
- \sum_{i} \frac{\partial}{\partial x_{\alpha i}} \bigg[M_{\alpha i}
(\vec{x}_\alpha,t)
P_\alpha(\vec{x}_\alpha,t)\bigg]
+ \frac{1}{2} \sum_{i, j} 
\frac{\partial}{\partial x_{\alpha i}}\frac{\partial}
{\partial x_{\alpha j}} \bigg[M_{\alpha i j}(\vec{x}_\alpha,t)
P_\alpha(\vec{x}_\alpha,t)\bigg]
\end{displaymath}
with
\begin{displaymath}
M_{\alpha i}(\vec{x}_\alpha,t) := m_{\alpha i}^{\alpha}(\vec{x}_\alpha,t)
+ \sum_{\beta} \int d^n x_\beta \, m_{\alpha i}^{\alpha \beta}
(\vec{x}_\alpha,\vec{x}_\beta,t)P_\beta(\vec{x}_\beta,t)  \, , 
\end{displaymath}
\begin{displaymath}
M_{\alpha i j}(\vec{x}_\alpha,t) := m_{\alpha i \alpha j}^{\alpha}
(\vec{x}_\alpha,t) + \sum_{\beta} 
\int d^n x_\beta \, \left[ m_{\alpha i \alpha j}^{\alpha \beta}
(\vec{x}_\alpha,\vec{x}_\beta,t)
+m_{\alpha i \beta j}^{\alpha \beta}
(\vec{x}_\alpha,\vec{x}_\beta,t) \right]
P_\beta(\vec{x}_\beta,t) 
\end{displaymath}
and
\begin{displaymath}
 \int d^n x_{\alpha} \, P_{\alpha}(\vec{x}_\alpha,t) = 1 \, .
\end{displaymath}
Here, the conventions
\begin{eqnarray*}
 m_{..}^\alpha (\vec{x}_\alpha,t) &\equiv& m_{..}^\alpha(\vec{X}^\alpha,t) \, ,
\nonumber \\
m_{..}^{\alpha\beta}(\vec{x}_{\alpha},\vec{x}_{\beta},t) &\equiv&
m_{..}^{\alpha\beta}(\vec{X}^{\alpha\beta},t) 
\end{eqnarray*}
have been used. 
\par
Distinguishing $A$ types $a$ of interaction only 
(see section \ref{Types}), we have
\begin{displaymath}
 P_\alpha \equiv P_a \quad \mbox{if} \quad 
\alpha \in a \, , 
\end{displaymath}
\begin{displaymath}
 m^\alpha_{..} \equiv \widetilde{m}^{a}_{..} \quad \mbox{if} 
\quad \alpha \in a \, , 
\end{displaymath}
\begin{displaymath}
 m^{\alpha\beta}_{..} \equiv \left\{ \begin{array}{ll}
\widetilde{m}^{ab}_{..} &\mbox{if } \alpha \ne \beta, \mbox{ }
 \alpha \in a, \mbox{ } \beta \in b \\
0 &\mbox{if } \alpha = \beta \, .
\end{array} \right.
\end{displaymath}
Therefore, we arrive at
\begin{equation}
 \frac{d}{dt}P_a(\vec{x},t) =
- \sum_{i} \frac{\partial}{\partial x_{i}} \bigg[M_{a i}
(\vec{x},t)
P_a(\vec{x},t)\bigg]
+ \frac{1}{2} \sum_{i, j} 
\frac{\partial}{\partial x_{i}}\frac{\partial}
{\partial x_{j}} \bigg[M_{a i j}(\vec{x},t)
P_a(\vec{x},t)\bigg] 
\label{Fok}
\end{equation}
with the {\em effective} drift coefficients
\begin{eqnarray}
M_{a i}(\vec{x},t) &:=& m_{a i}^{a}(\vec{x},t)
+ \sum_{b} \int d^n y \, m_{a i}^{ab}
(\vec{x},\vec{y},t)P_b(\vec{y},t) \label{ab1}\\
&=& \int d^n x' \, \Delta x'_i w^a(\vec{x}',\vec{x};t)\, , \nonumber
\end{eqnarray}
the {\em effective} diffusion coefficients
\begin{eqnarray}
M_{a i j}(\vec{x},t) &:=& m_{a i a j}^{a}
(\vec{x},t) + \sum_{b} 
\int d^n y \, \left[ m_{a i a j}^{ab}
(\vec{x},\vec{y},t)
+m_{a i b j}^{ab}
(\vec{x},\vec{y},t)\right]
P_b(\vec{y},t) 
\label{ab2} \\
&=& \int d^n x' \, \Delta x'_i \Delta x'_j
 w^a(\vec{x}',\vec{x};t) \nonumber \\
 &+& \sum_b \int d^n x' \! \int d^n y' \! \int d^n y \,
 \Delta x'_i \Delta y'_j w_{ab}(\vec{x}',\vec{y}';\vec{x},\vec{y};t) 
 P_b(\vec{y},t) \nonumber
\end{eqnarray}
and
%
\begin{eqnarray*}
 m^a_{..} &:=& \widetilde{m}^a_{..} \, , \\
 m^{ab}_{..} &:=& \left\{
\begin{array}{ll}
 N_b\cdot \widetilde{m}^{ab}_{..} & \mbox{if } b\ne a \\
 (N_a -1)\cdot \widetilde{m}^{ab}_{..}   
 & \mbox{if } b=a 
\,.
\end{array} \right. \nonumber
\end{eqnarray*}
The result (\ref{Fok}), (\ref{ab1}), (\ref{ab2})
could have also be obtained by a second order {\sc Taylor} 
expansion of the {\sc Boltzmann}-like
equations (\ref{Boltz1}), (\ref{abbrev})
(compare to \cite{Lib1,Montg,Paw}). Therefore,
the author suggests to term equations (\ref{Fok}), (\ref{ab1}), (\ref{ab2})
the ``{\sc Boltzmann-Fokker-Planck} equation''. 

\subsection{Selfconsistent solution} \label{SELF}

Equation (\ref{Fok}) is difficult to solve, since it is nonlinear in
$P_a(\vec{x},t)$ due to (\ref{ab1}), (\ref{ab2}). 
However, the stationary solution 
$P_a^{st}(\vec{x})$ of (\ref{Fok}) to (\ref{ab2}) can, with time independent
jump moments $m_{..}^a(\vec{x})$, $m_{..}^{ab}(\vec{x})$,
be obtained by recursive solution of the {\em linear}
differential equations
\begin{equation}
 0 = - \sum_{i} \frac{\partial}{\partial x_{i}} \bigg[M^k_{ai}
(\vec{x})
P_a^{k+1}(\vec{x})\bigg]
+ \frac{1}{2} \sum_{i, j} 
\frac{\partial}{\partial x_{i}}\frac{\partial}
{\partial x_{j}} \bigg[M^k_{a i j}(\vec{x})
P_a^{k+1}(\vec{x})\bigg] \, .
\label{recurs1}
\end{equation}
Here, we have used the definitions
\begin{eqnarray}
M^k_{ai}(\vec{x}) &:=& m_{a i}^{a}(\vec{x})
+ \sum_{b} \int d^n y \, m_{a i}^{ab}
(\vec{x},\vec{y})
P_b^k(\vec{y})  \, , 
\nonumber \\
M^k_{a i j}(\vec{x}) &:=& m_{a i a j}^{a}
(\vec{x}) + \sum_{b} 
\int d^n y \, \left[ m_{a i a j}^{ab}
(\vec{x},\vec{y})
+ m_{a i b j}^{ab}
(\vec{x},\vec{y})\right] P_b^k(\vec{y}) \, .
\label{recurs2}
\end{eqnarray}
Starting with suitable distributions $P^0_a(\vec{x})$, e.g.
the solutions of (\ref{Fok}) with neglected interaction terms
($m^{ab}_{..}(\vec{x},\vec{y}) \equiv 0$), we have
\begin{displaymath}
 P_a^{st}(\vec{x}) = \lim_{k \rightarrow \infty} P^k_a(\vec{x}) \, ,
\end{displaymath}
if the sequence of functions $P^k_a(\vec{x})$ converges. 
The procedure given by equations (\ref{recurs1}), (\ref{recurs2})
is completely
analogous to {\sc Hartree}'s method for calculating selfconsistent fields
in quantum mechanics \cite{Hartree,Fock}. 
In a slightly generalized form, it can also
be used to get dynamic solutions $P_a(\vec{x},t)$. A recursive solution
of {\sc Boltzmann}-like equations (\ref{Boltz1}), (\ref{abbrev}) or 
mean value equations (\ref{Mean}), (\ref{Abbrev})
is obtained in a similar way.

\section{Summary and fields of application}

Starting from a master equation for the state space, 
several equations for statistically 
behaving systems with pair interactions have been derived. For the case
of small (co)variances, approximate mean value and 
(co)variance equations have been
found. However, if the (co)variances are not negligible, {\sc Boltzmann}-like
equations are more appropriate.
A drastical reduction of the tremendous number of variables 
is obtained, if the systems can be divided into a
few types of interaction. Generalizations to 
interactions between an arbitrary number of systems have 
also been discussed.
\par
Important applications of these equations 
are not only known from thermodynamics \cite{Kampen1,Kubo,Keizer} 
and chemistry \cite{Kampen1,Opp,Chem1,Chem2,Chem3,Chem4,Chem5,Chem7},
but also from mathematical population
dynamics \cite{Weid1,Goel,Hallam,May} and from the biology of evolution 
\cite{Eigen,Ebeling}.
For example, the equation\\
\addtocounter{equation}{1}\setcounter{formel1}{\theequation}
\parbox{14.5cm}{\begin{eqnarray*}
 \frac{d}{dt}P(x,t) &=& \sum_{x'}
\bigg[w(x,x')P(x',t) - w(x',x)P(x,t)
\bigg]
\label{mutation} \\
&+& \bigg[f(x) - 
\sum_{y} f(y) P(y,t) 
\bigg] P(x,t) \label{selection}
\end{eqnarray*}}
\hfill \parbox{1.2cm}{\begin{displaymath}
\begin{array}{r}
\rule[-4mm]{0cm}{1.1cm}(\theequation{}\mbox{a})\\
\rule[-4mm]{0cm}{1.1cm}(\theequation{}\mbox{b})
\end{array}
\end{displaymath}}
describes mutations by the left-hand side of (\arabic{formel1}a), 
and a selection due to pair interactions 
by term (\arabic{formel1}b) \cite{Eigen,Ebeling}. It results from 
equations (\ref{Boltz1}), (\ref{abbrev}) with $n=1$, $A=1$, $w_1(x',x)
\equiv w(x',x)$ and
\begin{displaymath}
 w_{11}(x',y'; x,y;t)
\nonumber \\
:= \max\Big( f(x)-f(y),0 \Big)
\delta_{xx'}
\delta_{xy'}
+ \max\Big( f(y)-f(x),0 \Big)
\delta_{yx'}
\delta_{yy'} \, .
\end{displaymath}
$f(x)$ can be interpreted as the fitness of a species or strategy $x$ and
$\sum_{y} f(y) P(y,t) = \langle f \rangle$
as the {\em average} fitness.
\par
In addition, a master equation has been introduced allowing transition
rates which depend on the occupation numbers of the states. 
This equation has
important applications in quantitative social science 
\cite{Wei,Weid1,Weid2,Weid3,Weid4},
e.g. opinion formation \cite{Weid1,Hel1}.
\par
Finally, a {\sc Fokker-Planck} equation and a 
``{\sc Boltzmann-Fokker-Planck} equation'' for pair
interactions have been derived, since they will often be easier 
solvable or interpretable \cite{Hel2} than
the corresponding master equation resp. {\sc Boltzmann}-like equation.

{\bf Acknowledgements:} \\
The author wants to thank Prof. Dr. W. Weidlich 
for valuable discussions.

\end{document}